\documentclass[12pt]{iopart}
\pdfoutput=1
\usepackage{graphicx}
\expandafter\let\csname equation*\endcsname\relax
\expandafter\let\csname endequation*\endcsname\relax
\usepackage{amsmath}
\usepackage{amssymb}
\usepackage{bm}
\usepackage{cancel}
\usepackage[normalem]{ulem}
\usepackage{subfig}
\usepackage{placeins}
\usepackage{color}
\usepackage{iopams}

\usepackage{hyperref}

\newcommand{\sgn}{\text{sgn}}

\begin{document}
 
\title{Ripples in hexagonal lattices of atoms coupled to Glauber spins}
\author{M Ruiz-Garc\'{\i}a$^1$, L L Bonilla$^1$ and A Prados$^2$}

\address{$^1$Gregorio Mill\'an Institute for Fluid Dynamics, Nanoscience and Industrial Mathematics, Universidad Carlos III de Madrid,
Avenida de la Universidad 30, 28911 Legan\'es, Spain}
\address{$^2$F\'{\i}sica Te\'{o}rica, Universidad de Sevilla,
Apartado de Correos 1065, E-41080, Sevilla, Spain}

\begin{abstract}

  A system of atoms connected by harmonic springs to their nearest
  neighbors on a lattice is coupled to Ising spins that are in contact
  with a thermal bath and evolve under Glauber dynamics. Assuming a
  nearest-neighbor antiferromagnetic interaction between spins, we
  calculate analytically the equilibrium state{. On a one-dimensional
    lattice, the system exhibits} first and second order phase
  transitions. The order parameters are the total magnetization and
  the number of spin pairs in an antiferromagnetic configuration. On a
  hexagonal two dimensional lattice, spins interact with their
  nearest-neighbors and next-nearest-neighbors. Together with the
  coupling to atoms, these interactions produce a complex behavior
  that is displayed on a phase diagram. There are: ordered phases
  associated to ripples with atomic wavelength and antiferromagnetic
  order, ordered phases associated to ripples with nanometer
  wavelengths and ferromagnetic order, disordered glassy phases, and
  other phases presenting stripes formed by different domains. These
  static phases are discussed in relation to existing experiments and
  results for other models found in the literature.

\end{abstract}




\maketitle

\section{Introduction}

Graphene is a one-atom thick crystal membrane with extraordinary
mechanical and electronic properties
\cite{sci04nov,PNA05nov,RMP09cas}.  Experimental characterization of
suspended graphene shows that it is covered with ripples.  These
ripples are several nanometers long waves of the sheet without a
preferred direction \cite{nature07mey,pss09ban}, modify the electronic
band structure \cite{PRB08gui}, and are expected to be relevant in the
understanding of electronic transport in graphene
\cite{PTRS08kat}. Also, more recently, buckling in which the unit
structures consist of only two-three unit cells of the graphene
honeycomb lattice has been experimentally observed \cite{ASC11mao}.

There have been many studies of ripples.  The earliest studies using
Monte Carlo \cite{nat07fas} or molecular dynamics simulations
\cite{PRB07abe} have shown that ripples may be connected to variable
length $\sigma$ bonds of carbon atoms and may be caused by thermal
fluctuations. Other studies have explored the connection between
rippling and electron-phonon coupling \cite{EPL08Eun,PRB09gaz}. In
particular, it has been suggested that, at zero temperature, the
electron-phonon coupling may drive the graphene sheet into a quantum
critical point characterized by the vanishing of the bending rigidity
of the membrane \cite{PRL11san}. The continuation of this work by
J.~Gonzalez \cite{PRB14gon} discusses the nonzero expectation value of
the mean curvature (the Laplacian of the flexural phonon field) once
the bending rigidity of the membrane vanishes, and its role as order
parameter. Alternatively, assuming that the graphene membrane is
fluctuating in $2+d$ dimensions (with $d\gg 1$), Guinea \textit{et
  al.\/} have calculated the dressed two-particle propagators of the
elastic and electron interactions. They have found a collective mode
which becomes unstable at a nonzero wave vector and causes the
appearance of Gaussian curvature \cite{PRB14gui}. Amorim et al.\/
\cite{PRB14amo} estimate the crossover temperature between quantum and
classical descriptions to be 70-90 K. Thus a quantum description of
ripples is not necessary at room temperature. All these studies
investigate and characterize rippling as equilibrium phenomena.

We are interested in rippling dynamics and stability of static
corrugations under disturbances. In experiments to visualize ripples
using a transmission electron microscope (TEM), the suspended graphene
sheet is hit by a low-intensity electron beam that may push atoms
out-of-plane upward or downward in a random fashion. An alternative
technique to visualize ripples is using a scanning tunneling
microscope (STM)\cite{zan12}. In this case, the graphene sheet is
pushed and locally heated in the region close to the tip. Depending on
the tunneling current and the voltage between tip and sheet, the
latter may undergo a phase transition from a flexible (rippled) to a
rigid (buckled) state \cite{sch15}.

In Ref.~\cite{nl12oha}, the authors simplify the distortion
of the $2$d crystal by modeling it with two-state spin-like variables
($+1$ upward, $-1$ downward). There are antiferromagnetic interactions
among these spins, because the out-of-plane shift of the atoms in
opposite directions stabilize the strictly $2$d system while keeping
the gapless band structure of graphene \cite{nl12oha}. A rich phase
diagram is found, including paramagnetic, ordered and glassy phases,
depending on the temperature and the values of the nearest-neighbor
and next-nearest-neighbor couplings. In this way, they describe the
formation and origin of the atomic scale rippling found in
Ref.~\cite{ASC11mao}. On the other hand, there are also models that
investigate rippling by considering at each lattice site a continuous
variable $u$ describing the out-of-plane displacement of the carbon
atom coupled to a spin variable ($\sigma=\pm 1$) representing an
internal degree of freedom \cite{pre12bon,jsm12bon}. This may be
understood as a mechanical system coupled to a spin system. The spin
at each lattice site represents the non-saturated fourth bond that, by
a physical mechanism similar to the one discussed in \cite{nl12oha},
tries to pull the corresponding atom upward ($u>0$) or downward
($u<0$) from the flat sheet configuration. Mathematically, this is
done by introducing a linear coupling term in the system energy,
proportional to $-u\sigma$ for each lattice site.  In these simple
models, the mechanical system is either a chain of oscillators
\cite{pre12bon} or a discrete elasticity model of the hexagonal
graphene lattice \cite{jsm12bon}, while the spins are in contact with
a thermal bath and flip randomly according to Glauber dynamics at the
temperature $T$ of the thermal bath.  In both models, the system forms
metastable but long-lived ripples assuming slow spin relaxation
\cite{pre12bon,jsm12bon}.

It is worth investigating a combination of the two approaches
described in the previous paragraph. Firstly, it seems sensible to
model the out-of-plane displacement at each lattice site by a
continuous variable as in Refs.~\cite{pre12bon,jsm12bon}, which is
driven by the internal degree of freedom represented by the
spin. Secondly, these spins certainly interact among themselves, by
the mechanism proposed in \cite{nl12oha}. Thus, in this paper we
discuss the formation and dynamics of ripples in graphene through a
system of atoms connected by harmonic springs and coupled to
interacting Ising spins. We start from the spin-oscillator chain model
\cite{pre12bon,jsm12bon,jstat10} and add interactions among spins that
produce stable rippling states.  There appear different phases and
transitions between them depending on parameter values.

The plan of the paper is as follows. In Section \ref{model}, we
introduce the one-dimensional model and calculate analytically the
equilibrium state.  We also present the equations that determine the
dynamics of the system. Section \ref{2d} is devoted to the expansion
of the model to $2$ dimensions on a hexagonal lattice, with first and
second-neighbors interactions between spins. Besides, we do a
systematic analysis of the system modifying the parameters and
studying the stationary configuration after the transitory, using a
phase diagram in Section \ref{pd}, and describing the different
phenomenology of each phase in Section \ref{rc}. The main conclusions
are presented in Sec. \ref{con}. Relevant information that is not
covered in the main text is presented in the Appendices: some
geometrical expressions for the hexagonal lattice are discussed in
\ref{AC}, while images of the different phases are collected in
\ref{img}.

\section{The one-dimensional model}\label{model}

To start with, we consider a one-dimensional chain of $N$ oscillators
with nearest-neighbor interactions, in which each oscillator is
linearly coupled to an Ising spin $\sigma_i=\pm 1$. A detailed
investigation of this model can be found in
Ref.~\cite{pre12bon}. Therein, it was shown that, for appropriate
temperatures, the stable equilibrium configuration has only one
ripple, although there appeared some more complex long-lived
metastable rippled states. To explore whether stable multi-rippled
equilibrium configurations are possible, we add an anti-ferromagnetic
term to the hamiltonian,
\begin{equation}
 \mathcal{H}=\sum_{j=0}^{N} \biggl[ \frac{p_j^2}{2m}+ \frac{k}{2}(u_{j+1}-u_j)^2-fu_j\sigma_j 
 + J \sigma_{j+1}\sigma_j\biggr].
\label{H1}
\end{equation}
Here $u$ and $p$ are the vertical displacement and momentum
respectively, and the extreme oscillators and spins are fixed
($u_0=p_0=\sigma_0=u_{N+1}=p_{N+1}=\sigma_{N+1}=0$) \cite{bc}.
The dynamics {of the model} is  as follows: (i) the oscillators{'} equations of motion,
\begin{equation}
\label{eq:mot}
m\,\ddot{u}_j - k\, (u_{j+1}+u_{j-1}-2u_j)=f\sigma_j,
\end{equation}
{are the usual ones,} whereas the spins {evolve according to} Glauber dynamics \cite{Gl63}. The transition
rate from the configuration $(\bm{u},\bm{p},\bm{\sigma})$ to
$(\bm{u},\bm{p},R_j \bm{\sigma})$ (obtained from $\bm{\sigma}$ by
flipping the $j$-th spin) is
\begin{align}
W_j(\bm{\sigma}|\bm{u},\bm{p})=\frac{\alpha}{2}(1-\beta_j \sigma_j), \\ 
\beta_j= \tanh\! \left[\frac{f}{T} u_j -\frac{J}{T} (\sigma_{j-1}+\sigma_{j+1})\right]\!,
\end{align}
in which $\alpha$ is the characteristic attempt rate for the spin
flips. The Glauber transition rates ensure that the system
satisfies detailed balance, and therefore the system reaches
equilibrium for long enough times. In equilibrium, the probability of
a certain configuration $(\bm{u},\bm{p},\bm{\sigma})$ is proportional
to $e^{- \mathcal{H}/T}/Z$, where we measure the temperature $T$ in units of energy.

As explained in \cite{pre12bon}, for $J=0$ rippling appears
provided the temperature is less than
\begin{equation}
T_0=\frac{f^2K_{N}^{2}}{k}, \quad K_{N}\sim \frac{N}{\pi}, 
\end{equation}
as $N\to\infty$. To guarantee that the diffusive term in \eqref{eq:mot} remains finite in the continuum limit, it is convenient to nondimensionalize the equations of motion as follows: 
\begin{eqnarray}
&&
    u_j^*=\frac{k u_j}{f K_{N}^2},
   \quad t^*=\frac{t}{K_{N}}\sqrt{\frac{k}{m}},\label{nondim_u_t} \label{par1}\\
&& \kappa=\frac{J}{ T_{0}},\quad\delta=\frac{\alpha K_N\sqrt{m}}{\sqrt{k}},
   \quad 
   \theta=\frac{T}{T_{0}}=T \frac{k}{f^{2}K_{N}^{2}}.
\label{par2}
\end{eqnarray}
Then the transition rates and the equations of motion become
\begin{eqnarray}
&&W_j^*(\bm{\sigma}|\bm{u}^*,\bm{p})=\frac{\delta}{2}(1-\beta_j \sigma_j), \nonumber\\ 
&&\beta_j= \tanh\! \left[\frac{u^{*}_{j}}{\theta} -\frac{\kappa}{\theta} (\sigma_{j-1}+\sigma_{j+1})\right]\!,\nonumber\\
&&\frac{d^2u^*_j}{dt^{*\, 2}}-K_{N}^2(u^*_{j+1}+u^*_{j-1}-2u^*_j)= \sigma_j.  \label{nondim}
\end{eqnarray}
We will omit the asterisks in nondimensional equations from now on. In
order to obtain the scaling of the critical temperature, we need to
know the scaling of the model parameters with the system size. In
principle, this could be done by deriving our model from a fundamental
microscopic one, but this is outside the scope of this
paper. Nevertheless, we discuss some possible scalings in the
following. If both the elastic constant $k$ and the antiferromagnetic
coupling constant $J$ are considered to be independent of the system
size, the only remaining parameter is $f$, the coupling between the
elastic and internal (spin) degrees of freedom. If $f$ is also
independent of the system size, the critical temperature $T_{0}$
diverges as $N^{2}$. In this case, rippling would be observed at all
temperatures. On the other hand, a finite value of $T_{0}$ in the
large system size limit is obtained when $f\propto N^{-1}$. Then,
rippling would be observed only for $T<T_{0}$. In principle, both
situations are compatible with current experiments, in which rippling
is observed over a wide temperature range.

Let us consider now the equilibrium situation.  
Equation \eqref{nondim} can be averaged, with the result
\begin{equation}\label{curv}
 \frac{1}{\pi^{2}} \frac{d^{2}}{dx^{2}} \langle u\rangle=-\langle
 \sigma\rangle, 
\end{equation}
in which $\langle u\rangle$ and $\langle\sigma\rangle$ are the
equilibrium average height and spin at position $x=i/N$;
the system has unit size in the continuous space variable
  $x=i/N$, $0\leq x\leq 1$.  Therefore, the average curvature of the
ripples is directly linked to the average spin. Very recently, this
idea has been used to develop a phenomenological Ising model to study
rippling in graphene in scanning tunneling microscopy experiments, in
which each spin represents a whole ripple and the spin sign gives its
corresponding convexity \cite{sch15}. Interestingly, {we can derive} 
an  effective free energy for the string, by integrating
  the canonical distribution over the momenta $\bm{p}$ and the spins
  $\bm{\sigma}$: the resulting probability $\mathcal{P}$ becomes a
  functional of the string profile $u(x)$, which in dimensionless
  variables reads \cite{unpub}
\begin{subequations}\label{F-cont}
\begin{equation}
  \mathcal{P}[u] \propto \exp{\left(-\frac{\mathcal{F}}{\theta}\right)}, \quad 
  \mathcal{F}[u]=N \int_{0}^{1} dx \,
  \underbrace{\left[\frac{1}{2\pi^{2}}\left(\frac{\partial u}{\partial
        x}\right)^{2}-\theta \ln\zeta\left(
\frac{u}{\theta},\frac{\kappa}{\theta} \right)\right]}_{f(u,\frac{du}{dx})},
\end{equation}
\begin{equation}
\zeta\left(
\frac{u}{\theta},\frac{\kappa}{\theta} \right)=\exp\left(-\frac{\kappa}{\theta}\right)
         \cosh\left(\frac{u}{\theta}\right)+\exp\left(\frac{\kappa}{\theta}
\right)                \sqrt{1+\exp\left(-\frac{4\kappa}{\theta}\right)\sinh^{2}\left(\frac{u}{\theta}\right)}.
\end{equation}
\end{subequations}
The quantity $\ln\zeta$ is the logarithm of the spins
  partition function per site, which depends on the ``field''
  $f u_{j}/T \to u/\theta$ and the coupling $J/T\to
  \kappa/\theta$.
  The particularization of this free energy to the $J=0$ case was
  obtained in Ref.~\cite{pre12bon}. For $J\neq 0$, in order to
  calculate $\ln\zeta$, the system is divided into a set of nearly
  independent subsystems with $N_{s}\gg 1$ sites each, but such that
  $N_{s}\ll N$ and the ``field'' $u$ can be considered almost constant
  within each subsystem. We may denote the subsystems by $S(x)$, in
  which $x$ corresponds to the position of the subsystem in the
  continuum limit. In each subsystem, the local average magnetization
  $\mu=N_{s}^{-1}\sum_{j\in S(x)}\langle\sigma_{j}\rangle$ and
  correlation
  $C=N_{s}^{-1}\sum_{j\in S(x)}\langle\sigma_{j}\sigma_{j+1}\rangle$
  are {given by} the usual formulas
\begin{equation}
  \label{eq:magn_corr}
  \mu=\theta\frac{\partial\ln\zeta}{\partial u}, \quad C=-\theta\frac{\partial\ln\zeta}{\partial\kappa}.
\end{equation}
{as} $u$ plays the role of the external field and {$\kappa$} is the
  coupling constant. Global order parameters may be defined as
\begin{equation}
  \label{eq-global-or-param}
  M=\left|\,\int_{0}^{1} dx \, \mu \,\right|, \quad \mathcal{DL}=\frac{1}{2}\left(1-\int_{0}^{1} dx \, C\right),
\end{equation}
{in which $M$ is} the absolute value of the total magnetization
per site and the fraction of spin pairs in an antiferromagnetic
configuration {is $\mathcal{DL}=1$ } ($\mathcal{DL}=0$) for
perfect antiferromagnetic (ferromagnetic) ordering.

Equation~\eqref{F-cont} clearly shows that the
  free energy is an extensive quantity if the dimensionless variables
  are of the order of unity, which {is consistent with} our
  scalings.  Taking into account the expressions for the free energy,
  Eq.~\eqref{F-cont}, and the average magnetization,
  Eq.~\eqref{eq:magn_corr}, the equation giving the equilibrium
  profile \eqref{curv} is nothing but the Euler-Lagrange equation for
  the free energy functional, as it should be: the string profile
  minimizes the free energy. 
  
\begin{figure}
  \centering
\includegraphics[width=0.49\textwidth]{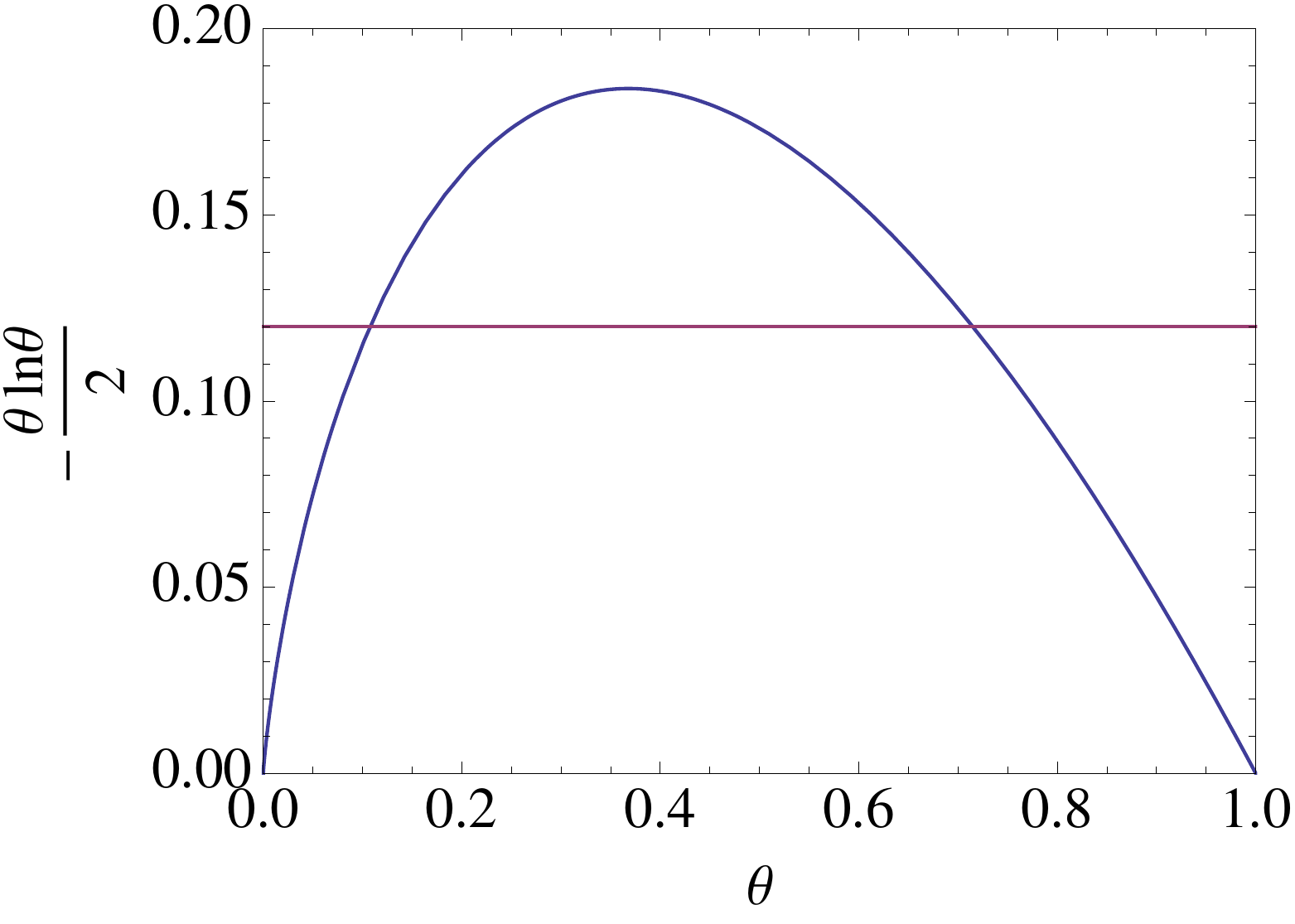}
\includegraphics[width=0.49\textwidth]{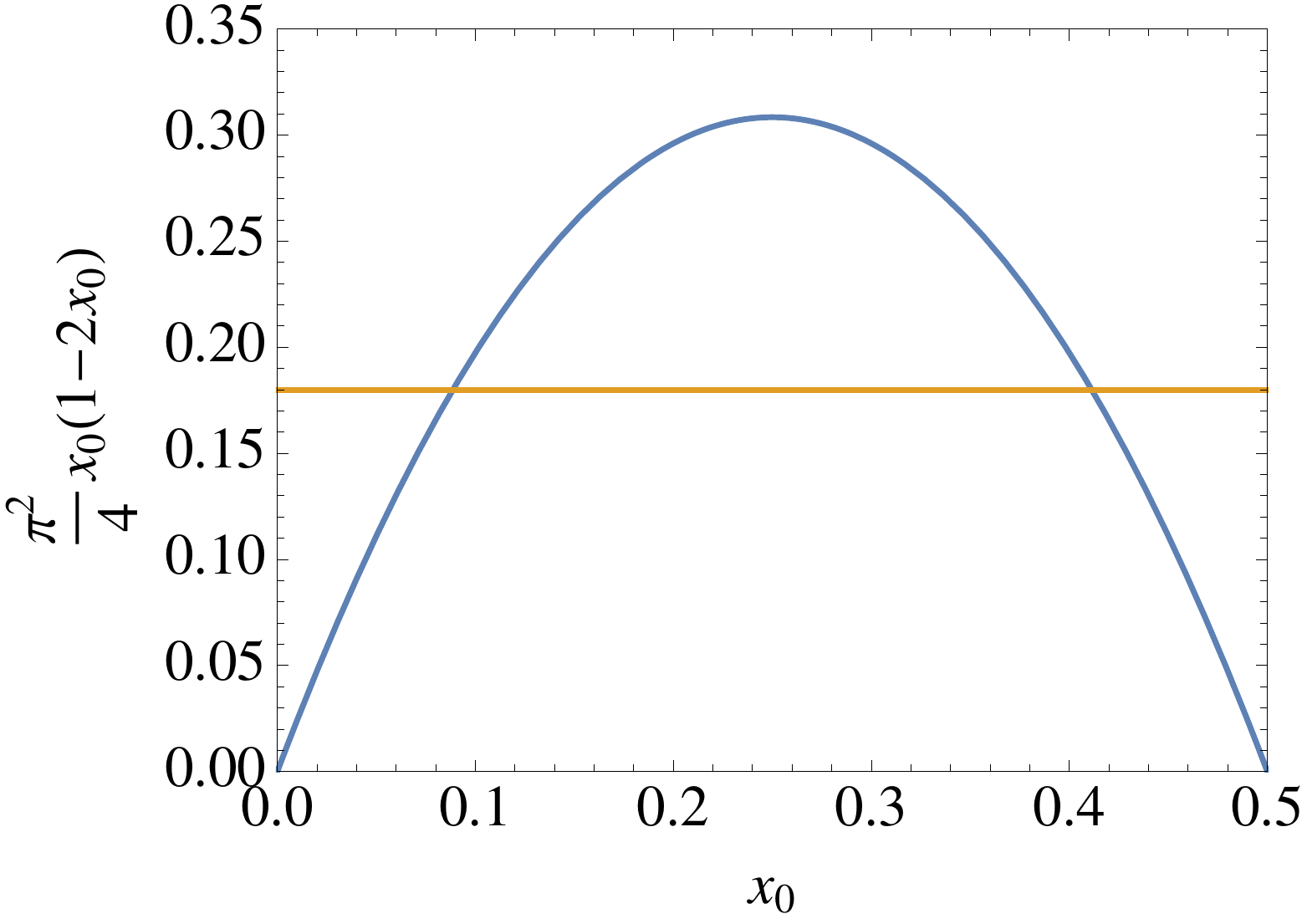}
\caption{\label{bif} (Left) Function controlling the bifurcation to the
  ferromagnetic phase. The bifurcation condition
  $\exp(-2\kappa/\theta)/\theta=1$ is equivalent to
  $y(\theta)=-\frac{1}{2}\theta\ln\theta=\kappa$. It is clearly seen
  that the bifurcation condition is fulfilled by two temperatures
  $\theta_{1}$ and $\theta_{2}$ for $\kappa$ smaller than the maximum $y_{\text{max}}=(2e)^{-1}$
  of $y$, that is, $\kappa<y_{\text{max}}$. (Right) Function controlling the
length $x_{0}$ of the antiferromagnetic regions near the borders of
the system at low temperatures. This length is given by
$\pi^{2}x_{0}(1-2x_{0})/4=\kappa$, which has two solutions $x_{01}$
  and $x_{02}$, $x_{02}=1/2-x_{01}$, for $\kappa<\pi^{2}/32\simeq 0.3$
  and no solutions for $\kappa>0.3$. At the limit value
  $\kappa=0.3$, it is $x_{01}=x_{02}=1/4$. }
\end{figure}

In what follows, we summarize the main physical implications
  of the short-ranged antiferromagnetic interaction, as compared to
  the $J=0$ case. The flat profile $u=0$ is always a solution of
  Eq.~\eqref{curv}, but it becomes unstable for
  $\exp(-2\kappa/\theta)/\theta>1$. For $J=0$, {there appear} rippled configurations
  with non-zero magnetization {that are stable} for
  $\theta<1$ \cite{pre12bon}{. F}or $J\neq 0$ the bifurcation
  condition $\exp(-2\kappa/\theta)/\theta=1$ {produces two temperatures}
  $\theta_{1}$ and $\theta_{2}$ for $\kappa<0.18$, as seen in the left
  panel of Fig.~\ref{bif}.  Specifically, these rippled configurations
  become unstable for low (high) enough temperatures,
  $\theta<\theta_{1}$ ($\theta>\theta_{2}$). The transitions at
  $\theta_{1}$ and $\theta_{2}$ are of second order, the order
  parameters $M$ and $\mathcal{DL}$ are continuous because $u$
  bifurcates continuously from the solution $u=0$, similarly to the
  behaviour found in the $J=0$ case. For $\kappa>0.18$, this rippled
  ferromagnetic phase no longer exists because the antiferromagnetic
  coupling is too strong. 

In the limit $\theta\to 0^{+}$, the partition function {of the spins becomes}
\begin{subequations}
  \label{eq:T=0}
\begin{equation}
\lim_{\theta\to 0}\theta\ln\zeta\left(
\frac{u}{\theta},\frac{\kappa}{\theta} \right)=\kappa+(|u|-2\kappa)\, \eta(|u|-2\kappa),
\end{equation}
\begin{equation}
\lim_{\theta\to 0}\mu= \sgn(u)\,\eta(|u|-2\kappa), \quad \lim_{\theta\to
0}C=\sgn(|u|-2\kappa).
\end{equation}
\end{subequations}
where $\eta(x)$ is the Heaviside step function, $\eta(x)=1$
  for $x>0$ and $\eta(x)=0$ for $x<0$, and $\sgn(x)$ is the sign
  function, $\sgn(x)=2\eta(x)-1$. In the flat configuration, $\mu=0$
  and $C=-1$ everywhere. {There appears a new rippled low temperature phase,
  that is antiferromagnetic near the boundaries because of} the clamped
  boundary conditions{. I}nside an interval of length
  $x_{0}$ close to the boundaries, $|u|<2\kappa$, and 
  $\mu=0$, $C=-1$ therein. Then, the simplest configuration is
  composed of (i) two straight lines ($u''=0$) near the boundaries,
  that is, in the intervals $(0,x_{0})$ and $(1-x_{0},1)$, and (ii) a
  parabolic ripple with $|u|>2\kappa$ ($u''=\pm 1$) in between, for
  $x\in(x_{0},1-x_{0})$, which corresponds to ferromagnetic ordering
  because $\mu= \pm 1$ and $C= 1$ for $|u|>2\kappa$. The continuity of
  $u$ and $u'$ at $x=x_{0}$ implies that
  $\pi^{2}x_{0}(1-2x_{0})=4\kappa$, which determines two possible
  values of $x_{0}$ for $\kappa<0.3$, as seen in the right panel of
  Fig.~\ref{bif}.  The configuration corresponding to the smallest
  value $x_{01}$ is the absolute minimum of the free energy for
  $0\leq x_{0}<1/8$, whereas both the flat string and the
  configuration corresponding to $x_{02}$ are metastable. For
  $1/4>x_{01}>1/8$, the absolute minimum of the free energy
  corresponds to the flat string, whereas both configurations
  corresponding to $x_{01}$ and $x_{02}$ are metastable. The
  transition at $x_{c}=1/8$ (which corresponds to $25$ percent of the
  spins being antiferromagnetic) is first order, because the order
  parameters $M$ and $\mathcal{DL}$ change discontinuously: in the
  zero temperature flat configuration, it is $M=0$ and
  $\mathcal{DL}=1$, whereas in the rippled state we have $M=1-2x_{0}$
  and $\mathcal{DL}=2x_{0}$. Moreover, as is usually the case in
  first-order phase transitions, the string has many other metastable
  configurations: they have $n$ internal nodes $x_{i}$,
  $i=1,\ldots,n$, at which $u$ changes sign. The existence of the
  different phases and their relative stability will be thoroughly
  discussed elsewhere \cite{unpub}.

\section{The two-dimensional model} \label{2d}

\begin{figure}
\centering
\includegraphics[width=0.6\textwidth]{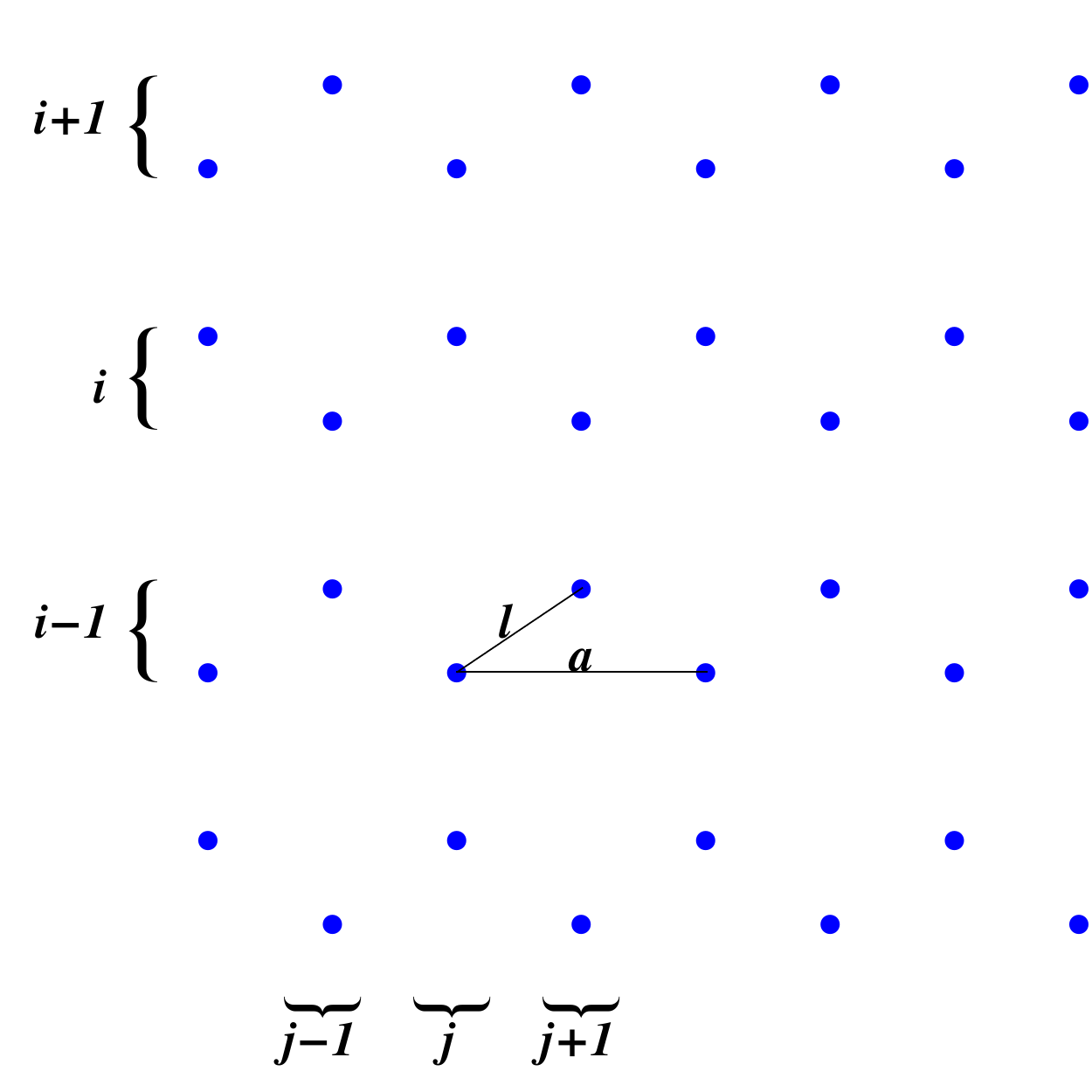}
\caption{Figure summarizing the atoms indexes and the
  parameters of the unit cell of the hexagonal lattice.}
\label{esq}
\end{figure}

Here, we extend the model to dimension $d=2$, in the
hope that this will make it possible to find more complex behaviors.
This extension is almost direct and, as we are
interested in applying the model to mimic a graphene sheet, we
consider a hexagonal lattice. Due to the symmetry, it is important to
write the hamiltonian carefully. First, each atom is indexed: $i$ will
be the row index and $j$ the column index, with the peculiarity that
each row comprises atoms with two different heights in a zigzag
distribution, see Figure \ref{esq}.  It is important to note that the
form of the equations will be qualitatively different for atoms for
which $|i-j|$ is an even number \textit{(e-atoms)}, which have one
nearest neighbor above and two below, and those for which $|i-j|$ is
an odd number \textit{(o-atoms)}, which have one nearest neighbor
below and two above, that is, the opposite situation. It is quite
obvious that if the plane is rotated by an angle of $\pi$, the two
types of atoms are interchanged.

Taking into account the notation described above, we can write down
the extension of the 1$d$ Hamiltonian to $d=2$. Moreover, we introduce
next-nearest-neighbor interactions,
\begin{eqnarray}
\mathcal{H} &=&\sum_{ij} \left[\frac{p_{ij}^2}{2m}-fu_{ij}\sigma_{ij}
  + J'
  \sigma_{ij}(\sigma_{i-1,j-1}+\sigma_{i,j-2}+\sigma_{i+1,j-1})\right]
  \nonumber
 \\ &+& \sum_{|i-j|=\text{even}} \Bigg\{\frac{k}{2}\left[
  (u_{ij}-u_{i+1,j})^2+
  (u_{ij}-u_{i,j-1})^2+(u_{ij}-u_{i,j+1})^2\right] \nonumber \\
  && \qquad\qquad+ J
  \sigma_{ij}(\sigma_{i+1,j}+\sigma_{i,j-1}+\sigma_{i,j+1})\Bigg\}, 
\label{H}
\end{eqnarray}
where $i$ and $j$ take values $1 \to i_{\text{max}}$ and $1 \to
j_{\text{max}}$, respectively.
Following the same steps as in the previous section, the
nondimensional equation of motion for each atom and the expressions
for the transition rate become
\begin{equation}
 \ddot{u}_{ij} - K_N^2 (u_{i+1,j}+u_{i,j-1}+u_{i,j+1}-3u_{ij})= \sigma_{ij}, \label{mov}
\end{equation}
\begin{equation}
\omega_{ij}(\bm{\sigma}|\bm{u})=\frac{\delta}{2}(1-\gamma_{ij}
\sigma_{ij}),
\end{equation}
\begin{eqnarray}
\gamma_{ij}&=&\tanh[\frac{u_{ij}}{\theta} -\frac{\kappa}{\theta}(\sigma_{i+1,j}+\sigma_{i,j+1}+\sigma_{i,j-1})\nonumber
\\ &&  \;\;
      -\frac{\lambda}{\theta}(\sigma_{i,j-2}+\sigma_{i,j+2}+\sigma_{i-1,j-1}
+\sigma_{i-1,j+1}+\sigma_{i+1,j-1}+\sigma_{i+1,j+1})],
\label{gamma}
\end{eqnarray}
where $K_N$ is a large scale parameter to be calculated later. As we
said before, the difference between e-atoms with o-atoms follows from
the rotation by $\pi$ of the plane. For that reason, only equations
for e-atoms have been written. In the latter equations, the height
variable $u$ and time are dimensionless.  In the
nondimensionalization, the same parameters as in equations
\eqref{par1} and \eqref{par2} appear, with the addition of
$\lambda=J'/T_{0}$, which corresponds to the new next-nearest-neighbor
interaction.

The length of each side of the finite hexagonal lattice is
$\tilde{L}=[3(n-1)+1]\tilde{l}/2$, where $\tilde{l}$ is the side of a
unit hexagonal cell and $n$ is the maximum value of the row index $i$
in Fig.\ \ref{esq}. Let us measure all lengths in units of
$\tilde{L}$, so that $l=\tilde{l}/\tilde{L}$ tends to zero as the
hexagonal lattice fills the plane. Then the expression within
parenthesis in \eqref{mov} has the limit
\cite{prb08car,prb12bon86,jsm12bon}
\begin{equation}
 u_{i+1,j}+u_{i,j-1}+u_{i,j+1}-3u_{ij} \rightarrow  \frac{a^2}{4}(\partial_x^2 u +\partial_y^2 u), \label{lap}
\end{equation}
as $a=\sqrt{3}l\to 0$. 
Therefore, we take $K_{N}$ proportional to $a^{-1}$, namely
\begin{equation}
K_N=\frac{\sqrt{2}}{\pi} a^{-1}=\frac{3n-2}{\sqrt{6}\pi}\propto n,
\end{equation}
to guarantee that the diffusive term in \eqref{mov} remains finite as
$l\to 0$ (continuum limit). Note that the increments of the
  continuous variables are $\Delta x(i\to i+1)=3l/2$ and
  $\Delta y(j\to j+1)=a/2=\sqrt{3}l/2$, as seen in Fig.~\ref{esq}, so
  that the hexagonal lattice goes to the unit square $0\leq x,y \leq
  1$ in the continuum limit. For details, see \ref{AC}.

Once the system reaches the stationary state, equation \eqref{mov} can be averaged ignoring thermal fluctuations. Thus, using equation \eqref{lap} for $n \gg 1 $ we get
\begin{equation}
\label{cont_2d}
\frac{1}{2\pi^{2}}\nabla^2 \langle u\rangle=- \langle \sigma \rangle,
\end{equation}
where $\langle u\rangle$ and $\langle \sigma\rangle$ are the
average height and spin at the point $(x,y)$ of the
  unit square. For $\kappa=\lambda=0$, we have that
  $\langle\sigma\rangle=\tanh(u/\theta)$ and the flat configuration
  $\langle u\rangle=0$ becomes unstable at $\theta=1$, similarly to
  the situation in the 1d case. This kind of rigidly buckled
  configurations have been observed in graphene in recent STM
  experiments \cite{sch15}. Equation \eqref{cont_2d} tells us that there is a correspondence between lattice
patterns given by the average height profile and the spin
configuration. Specifically, the curvature of the rippling is
  directly proportional to the average spin. Therefore, in the
following section we will mainly characterize the phases by the spin
configuration. 

\subsection{Phase diagram} \label{pd}

{As it has already been said, the stable steady state is a rigidly buckled configuration below the critical temperature ($\theta<1$), provided there is no interaction between spins}. We expect that the introduction of the nearest neighbour interaction among the spins should introduce new phases{. By analogy with the 1d system, an antiferromagnetic nearest neighbour
  interaction should make antiferromagnetic ordered phases
  to appear} for low enough temperatures. Looking for a more
complex phenomenology, we introduce a next-nearest-neighbor
interaction, as in Ref.~\cite{nl12oha}. This term appears in \eqref{H}
through $J'$ and in \eqref{gamma} through its dimensionless
counterpart $\lambda$. 

It is important to note that both $J$ ($\kappa$) and $J'$
($\lambda$) may take positive or negative values, corresponding to
antiferromagnetic and ferromagnetic interactions,
respectively. However, only positive values of $\lambda$ will be
considered, since a next-nearest-neighbor ferromagnetic interaction
just strengthens the nearest-neighbor one \cite{note}.  For positive
values of $\kappa$ and $\lambda$ the qualitative behavior is quite
different. The nearest-neighbor interaction provides a defined minimum
energy distribution in which each spin and its nearest-neighbors point
in opposite directions. However, the next-nearest-neighbor interaction
does not yield a defined minimum energy distribution. In fact, the
second-neighbors of each atom are second-neighbors to each other, and
therefore the next-nearest-neighbor interaction causes the system to
be frustrated \cite{nl12oha}. An enlightening discussion about
frustration is given in the introduction of \cite{mezard}. In
  principle, it is tempting to exclude negative values of $\kappa$
  from the analysis. On intuitive grounds, one may conclude that the
  nearest-neighbour ferromagnetic coupling with $\kappa<0$ should only
  strengthen the already long-ranged ferromagnetic interaction among
  the spins induced by the spin-lattice coupling term
  $-f u_{ij}\sigma_{ij}$ \cite{pre12bon}. Nevertheless, the situation
  is a little bit more complex, as discussed below.

\begin{figure}
\centering
\includegraphics[width=0.6\textwidth]{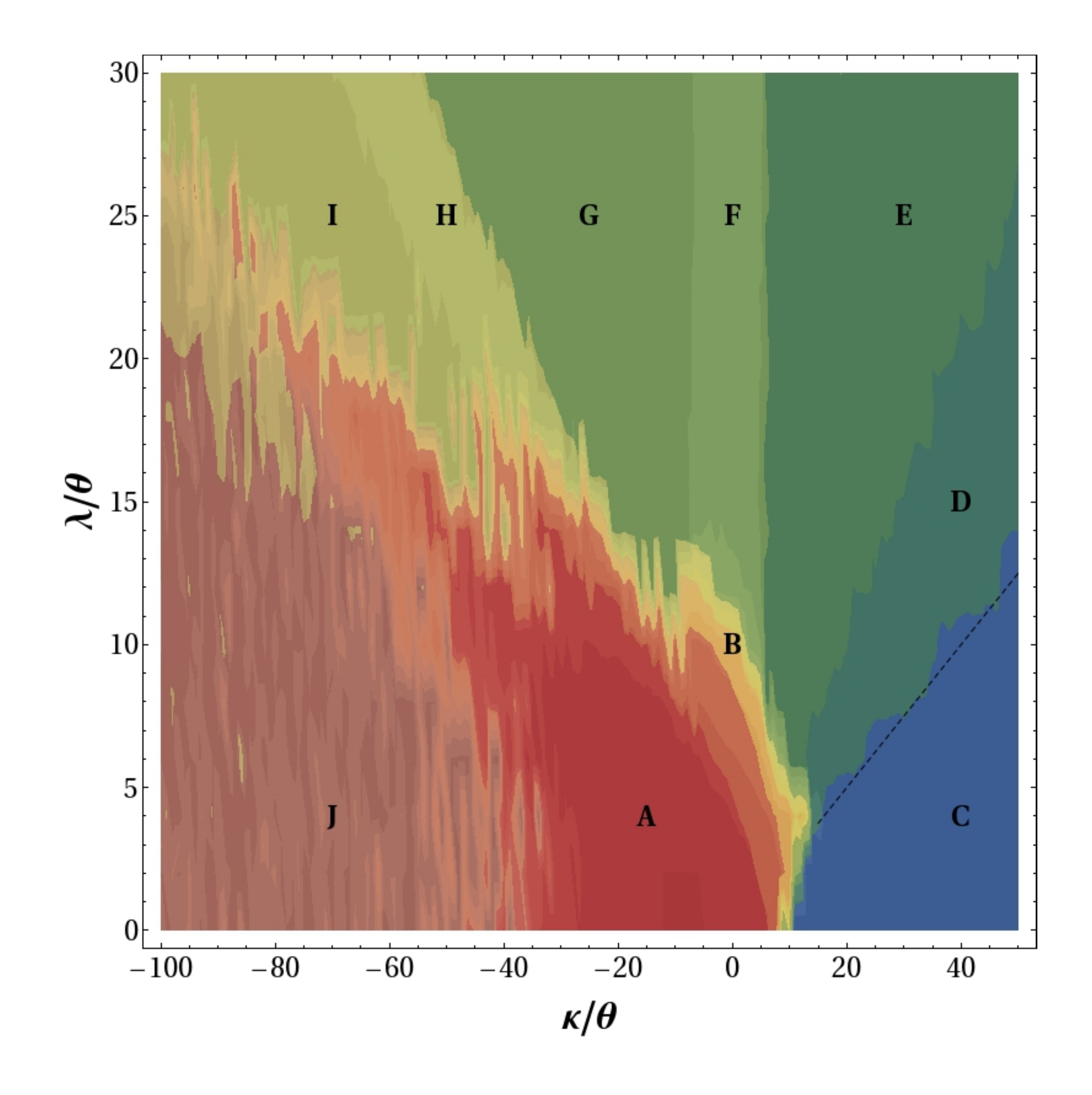}
\caption{Phase diagram for a hexagonal lattice coupled to Ising
  spins. Different regions have been delimited using the domain-wall
  parameter, the magnetization and the specific heat. Once
  the equilibrium state is reached, each region has a
  different behavior, which is explained in the
  text. Also plotted is the line $\kappa/\lambda=4$, which is
    a good estimate for the transition line between zones $C$ and $D$.
    This agrees with the line separating phases Ordered $1$ and $2$ in
    Ref.~\cite{nl12oha}.}
 \label{diag}
\end{figure}

We plot a phase diagram to show in only one graph all the different
behaviors, see Figure \ref{diag}. In our simulations, we have chosen
{a nondimensional temperature $\theta=0.01$,} which is far
below critical for $\kappa=\lambda=0$. A key parameter is
\begin{equation}
\mathcal{DL}=\frac{1}{N}\sum_{|i-j|=even} [3+\sigma_{ij}(\sigma_{i+1,j}+\sigma_{i,j-1}+\sigma_{i,j+1})],
\end{equation}
where $N$ denotes the number of atoms in the lattice. 
This parameter estimates the domain-wall length
\cite{nl12oha}, and it is equal to $3$ (resp.~$0$) for
completely ferromagnetic (resp.~antiferromagnetic) behavior. In
addition, to delimit the regions on the diagram, we have used the
absolute value of the usual magnetization
\begin{equation}
M=\left|\frac{1}{N}\sum_{ij}\sigma_{ij}\right|\!,
\end{equation}
and energy fluctuations (proportional to the specific heat), 
\begin{equation}
F=\sqrt{\langle(\Delta^{*}
e)^{2}\rangle}.
\end{equation}
Here $e$ is the system energy and $\Delta^{*}e=(e-\langle e\rangle)$,
where the angular brackets stands for the mean value that is calculated once the stationary state has been reached.

\subsection{Region characterization} \label{rc}

The different regions in the phase diagram have been characterized
using the three parameters $\mathcal{DL}$, $M$ and $F$. Figure
\ref{diag} is the superposition of the projections of $M$ and
$\mathcal{DL}$ on the plane $\lambda/\theta-\kappa/\theta$. Each region of the plane
correspond to different combinations of $M$ and $\mathcal{DL}$ values.
Once the regions have been delimited using the magnitudes described
above, the system is allowed to evolve with $\kappa$ and $\lambda$ in
one of the regions. Next, we verify that the system reaches the
equilibrium state and we obtain the basic structures in the spin
domains. Moreover, to check that we have actually reached the
equilibrium state, another simulation is carried out with this
distribution as the initial condition: if, aside from thermal
fluctuations, no evolution is found, equilibrium has been reached.

\begin{itemize}
\item Region $A$. $\mathcal{DL}\sim 3$, $M\sim 1$. The plane is
  completely curved, and the spins are all pointing in the same
  direction. This situation corresponds to small values of
    $\kappa$ { and $\lambda$}, for which the interaction that dominates is the one
  between the surface and the spins, in agreement with the simple
  picture already present in the 1$d$ model, see Section \ref{model}.
\item Region $B$. It is the zone surrounding $A$, on which
  $\mathcal{DL}$ and $M$ decrease from the $A$ values to those on the
  other regions. Here, the system displays a behavior that is
  analogous to the one described at the end of Section \ref{model} (in
  1$d$). The interaction between the surface and the spins is an
  effective ferromagnetic interaction with an intensity that decreases
  from the center to the border. Thus, the plane is curved but the
  spins close to the border are antiferromagnetically arranged.
\item Region $C$. $\mathcal{DL}\sim 0.5$, $M\sim 0$. The predominant
  interaction is the antiferromagnetic first-neighbor one. The
  equilibrium state (starting from a random initial spin distribution)
  is composed of antiferromagnetic domains.
\item Region $D$. $\mathcal{DL}$ increases from $0.5$ to $1.2$, $M\sim
  0$. The states in this region are metastable. Taking the
  distribution corresponding to the equilibrium state of $C$ or $E$ as
  the initial condition, the system does not evolve to the other
  state, at least in a simulation time much greater than the
  relaxation time from random initial conditions.
\item Region $E$. $\mathcal{DL}\sim 1.2$, $M\sim 0$. In this case, the
  spins are distributed in rows of two atoms in the lowest energy
  configuration. Beginning with random initial conditions, these
  two-atoms domains were created, with the rows in any of the three
  symmetrical directions.
\item Region $F$. $\mathcal{DL}\sim 1.5$, $M\sim 0.2$. The interaction
  between the plane and the spins is relevant again, since the
  antiferromagnetic next-nearest-neighbor interaction (with $\kappa
  \sim 0$) has no defined minimum energy distribution.  The plane is
  curved, leading to a non-zero magnetization.
\item Region $G$. $\mathcal{DL}\sim 1.8$, $M\sim 0$. The typical
  equilibrium configurations are long serpentine lines, with zero
  magnetization. Taking as initial conditions the spins arranged in
  rows, the system remains static.
\item Regions $H$ and $I$. In them, the system evolves from the
  characteristic configurations of $G$ to the ferromagnetic
  configurations of $J$, with the difference that in $H$ the
  magnetization is different from zero whereas in $I$ it is not.  $I$
  is a ferromagnetic first-neighbor state, but with domains smaller
  than in $J$ ($\mathcal{DL}$ smaller than in $J$).
\item Region $J$. $\mathcal{DL}\sim 2.5$, $M\sim 0.1$. In this case
  the system behaves as a completely ferromagnetic first-neighbor
  system. Starting from random initial conditions, ferromagnetic
  domains grow until reaching a stationary state. In this case $M$ is
  close to zero since spins are pointing to different directions in
  adjoining domains.
\end{itemize}
The plots of the typical equilibrium configurations for each region
are in \ref{img}.

It should be noted that our phase diagram does not contain a
paramagnetic state because the chosen temperature, $\theta=0.01$, is
far below the critical temperature for $\kappa=0$ and
  $\lambda=0$ (unity in our dimensionless variables). Each point of
the phase diagram corresponds to the energy minimum to which the
system evolves for the considered parameters. Once it is in the
neighborhood of this minimum, the energy barriers are so high that
ergodicity is no longer valid, and the system remains \textit{frozen}
\cite{hemmen}. This causes an Edwards-Anderson order parameter
\cite{edw},
 \begin{equation}
  q_{EA} \equiv \frac{1}{N} \sum_{ij} \mu_{ij}^{2}, \quad \mu_{ij} \equiv  \langle \sigma_{ij} \rangle,
  \label{q}
 \end{equation}
to be different from zero at every point of the
  plotted phase diagram. On the other hand, close to
$\kappa=\lambda=0$, the order parameter $q_{EA}$ will vanish as
  $\theta\to \infty$, once ergodicity is recovered. In
Eq.~\eqref{q} the average should be understood as a time average or an extended Gibbs average in a phase space composed of disjoint ergodic components \cite{hemmen}.

 \section{Conclusions} 
 \label{con}

 We have studied a system of atoms connected by harmonic springs and
 coupled to Glauber spins. The spins are in contact with a thermal
 bath and interact with their neighbors. The $1d$ system forms one
 ripple and becomes antiferromagnetic at the boundaries as $\rho$
 increases, until it becomes completely antiferromagnetic. When the
 system is on a $2d$ hexagonal lattice, each spin interacts with its
 nearest-neighbors and next-nearest-neighbors, aside from the coupling
 with the out-of-plane displacement. This situation generates
 different phases which are included in a phase diagram.

 The range of parameters in our phase diagram includes negative values
 for the nearest neighbor coupling constant $\kappa$ and is thus wider
 than the one used in \cite{nl12oha}, in which only antiferromagnetic
 interactions were considered. The change in the sign of the spin-spin
 interaction can be produced by the scattering of the conduction
 electrons at the spins, see \cite{ruder,binder}. We are interested in
 zero magnetization phases since they correspond to no overall
 bending. Our model provides different phases obeying this constraint:
 I and J are long wave length phases, similar to those observed in
 \cite{nature07mey,pss09ban}, whereas C, E and G are phases with
 atomic wave length. G is a stripy phase (see Figure \ref{FG}),
 which could be associated with patterns seen in
 \cite{ASC11mao}. The atomic wave length phases C and E correspond
 with phases Ordered 1 and 2, respectively, from
 Ref. \cite{nl12oha}. Therein, the line between these two phases is
 (in our variables) $\kappa/\lambda=4$, which agrees with the limit of
 true stability of C here. Interestingly, neither the metastable phase
 $D$ or the other phases (including the long wavelength phases I and
 J) were found in Ref.~\cite{nl12oha}. In that reference, (i) only
 positive values of $\kappa$ were considered, and (ii) there was no
 spin-atom coupling. 
 
The buckling phase A is surrounded by rippled phases C, E, and G with
no overall bending.  Starting from a point of the phase
  diagram belonging to region C (rippled phase), if we
increase the temperature while keeping $\kappa$ and $\lambda$
  (supposed temperature independent)
  fixed, we move along a straight line of slope $\lambda/\kappa$ in
  Figure \ref{diag} from phase C to the buckled phase A.  In
experiments with STM at fixed current, the temperature is locally
increased at the tip region and this triggers a transition from a
rippled flexible phase to a rigid buckled phase \cite{sch15}. Thus our
model contains the ripples-to-buckling transition observed in
experiments although more work needs to be done to explain STM
observations in detail \cite{unpub}.

To conclude, our model is based in a few parameters controlling simple interactions
which generate complex collective behaviors. This allows us to
identify the interactions responsible for each pattern.  In addition,
the elastic feature of the model makes it possible to visualize and
quantify the magnitude of the rippling, which could be compared with
experiments once height measurements had been improved.

\ack
This work has been supported by the Spanish Mi\-nisterio de Econom\'\i
a y Competitividad grants FIS2011-28838-C02-01 (MRG \& LLB), and
FIS2011-24460 (AP). MRG acknowledges  support from Ministerio de
Educaci\'on, Cultura y Deporte through the FPU program grant FPU13/02971.

\appendix

\section{Geometrical expressions}\label{AC}

We want our hexagonal lattice to have equal overall length and height. Let $n\equiv i_{max}$ be the total number of rows. Then
\begin{equation}
j_{max}=\text{IntegerPart}\left[\frac{3 (n-1)+1}{\sqrt{3}}+1\right],
\end{equation}
is the total number of columns, and the height of the hexagonal lattice is 
\begin{equation}
\tilde{L}=\frac{3 (n-1)+1}{2}\tilde{l},
\end{equation}
where $\tilde{l}$ is the length of the side of a unit hexagonal
cell. With these expressions, if $n=25$, then $j_{max}=43$ and the
vertical and horizontal side of the lattice are $1$ and $0.997$
respectively, in units of $\tilde{L}$. Our finite hexagonal lattice is
then roughly inscribed in a square and the nondimensional side of the
unit hexagonal cell is $l=\tilde{l}/\tilde{L}=0.027$.

\section{Phase diagram images}
\label{img}

In our simulations, we have used a lattice of 2,150 atoms and a
temperature $\theta=0.01$. We need to impose initial
and boundary conditions for the membrane and the spins. Initially, the
spins are in a completely random configuration, whereas the membrane
is flat and at rest. The membrane is clamped (zero displacement) at
the boundaries. As nearest and next-nearest neighbors determine the
dynamics of a given spin, see eq.~\eqref{gamma}, a spin located
next to the boundary condition needs data from nearby spins
located at the boundaries and also outside the lattice. The
simplest possibility is that the spins of clamped boundary
atoms and their nearest neighbors outside the lattice do not
interact with the others, which can be achieved by formally
assigning spin zero to them.
 
\begin{figure}[h!]
\centering
A
\includegraphics[width=2.75in]{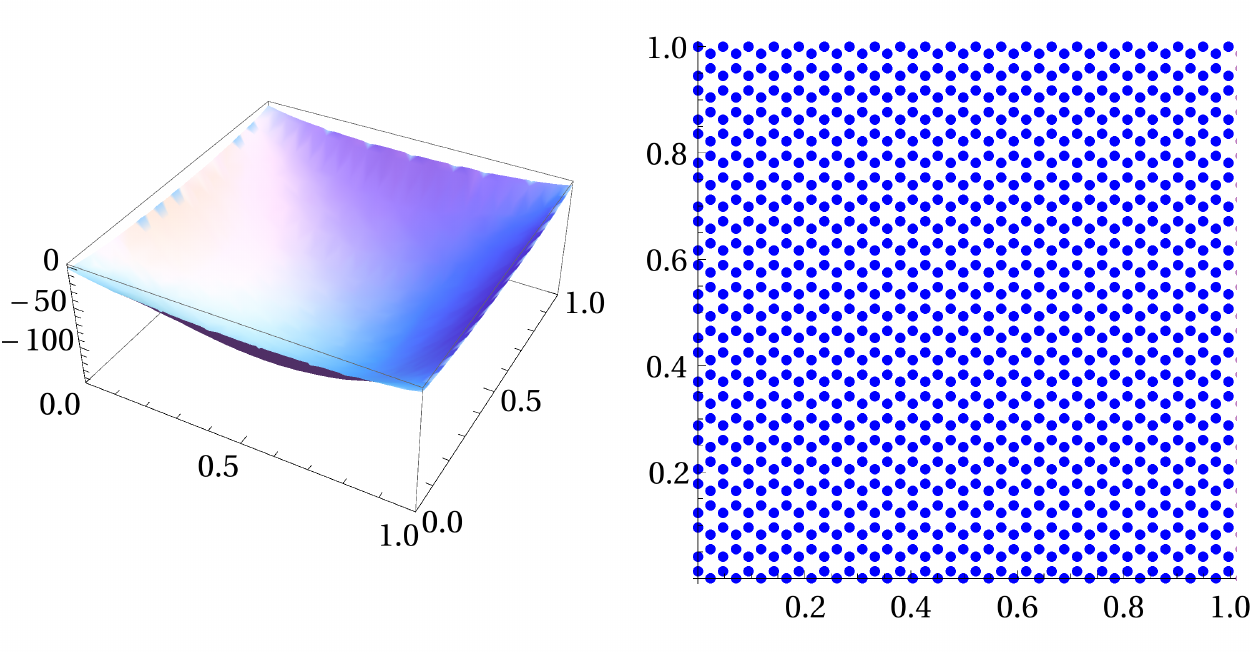}
\hfill
B
\includegraphics[width=2.75in]{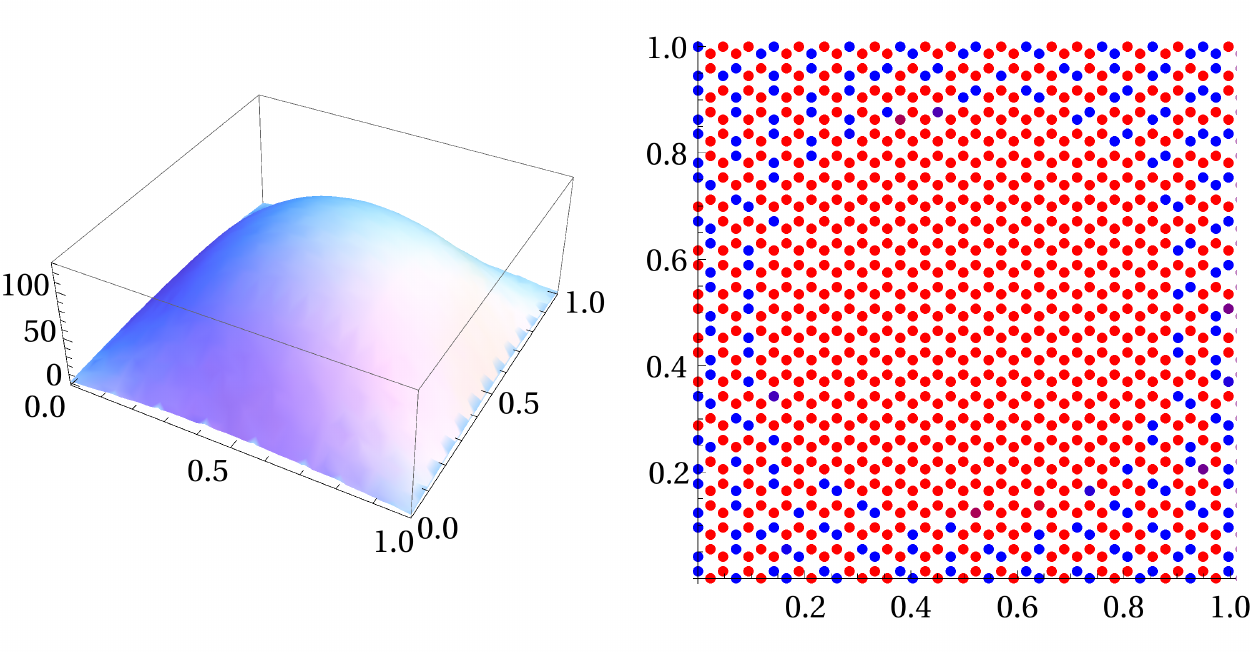}\\
C
\includegraphics[width=2.75in]{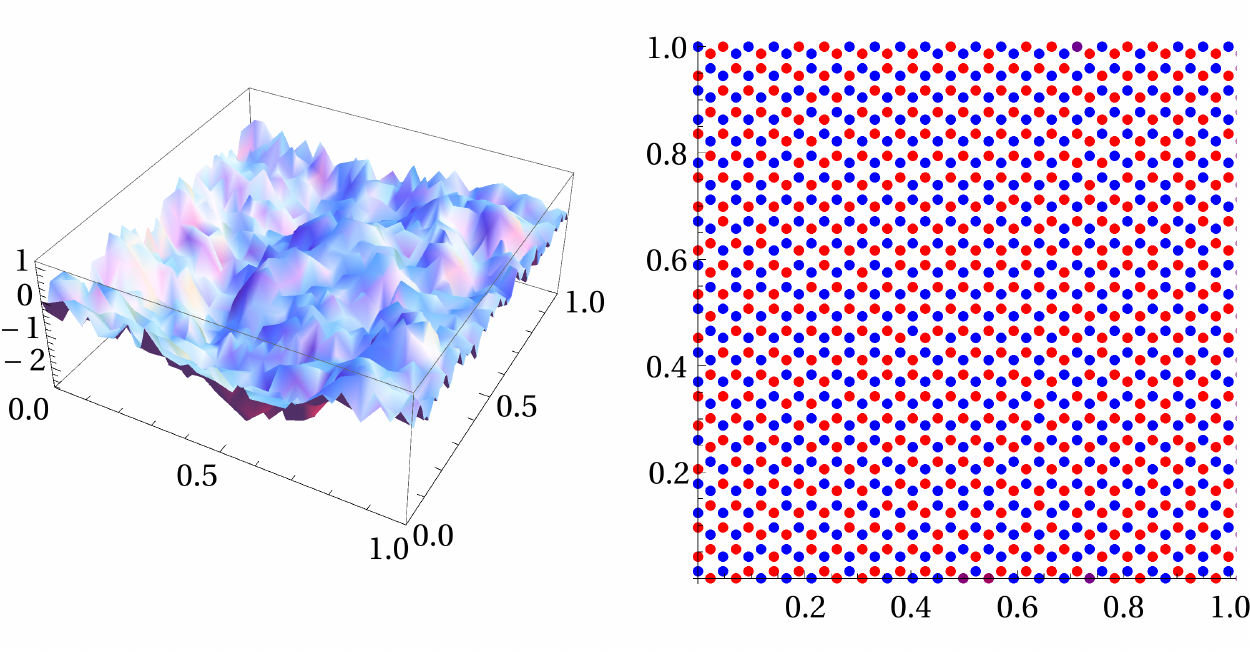}
\hfill
E
\includegraphics[width=2.75in]{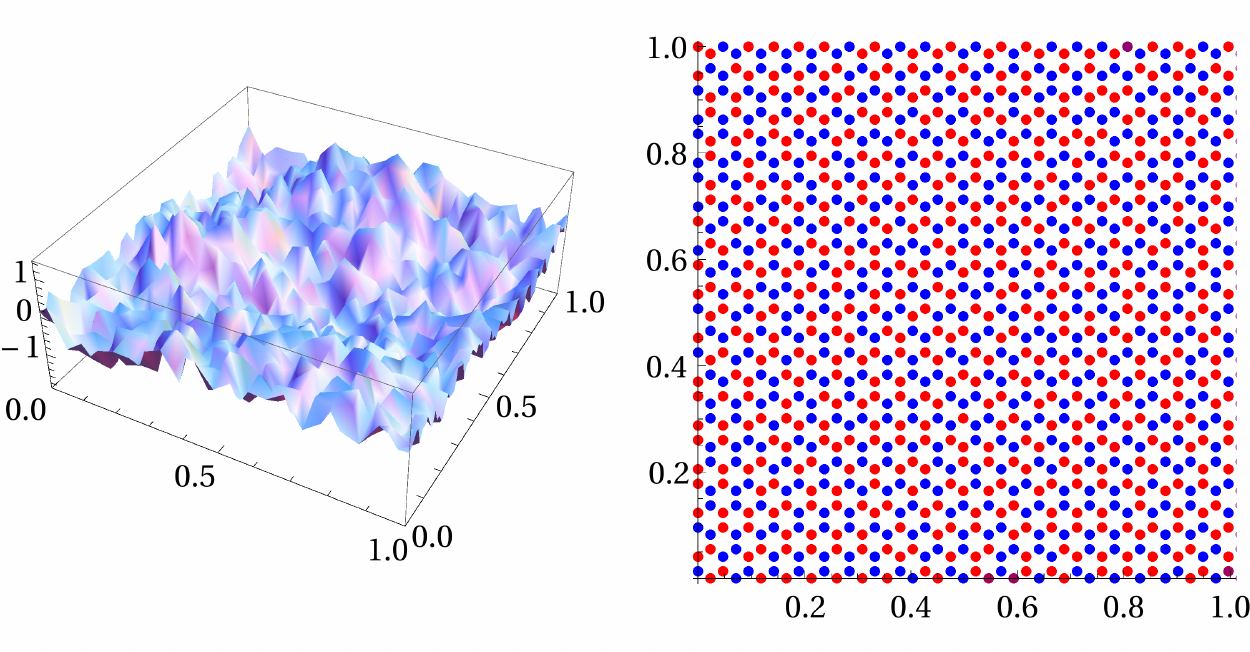}\\
F
\includegraphics[width=2.75in]{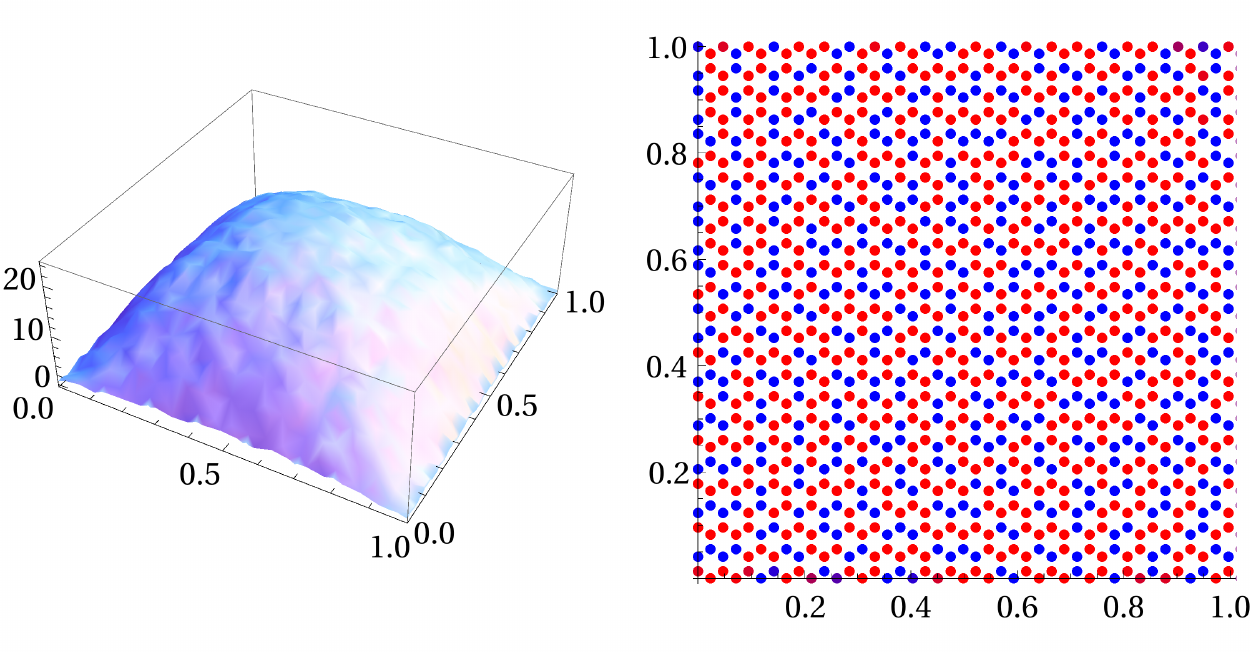}
\hfill
G
\includegraphics[width=2.75in]{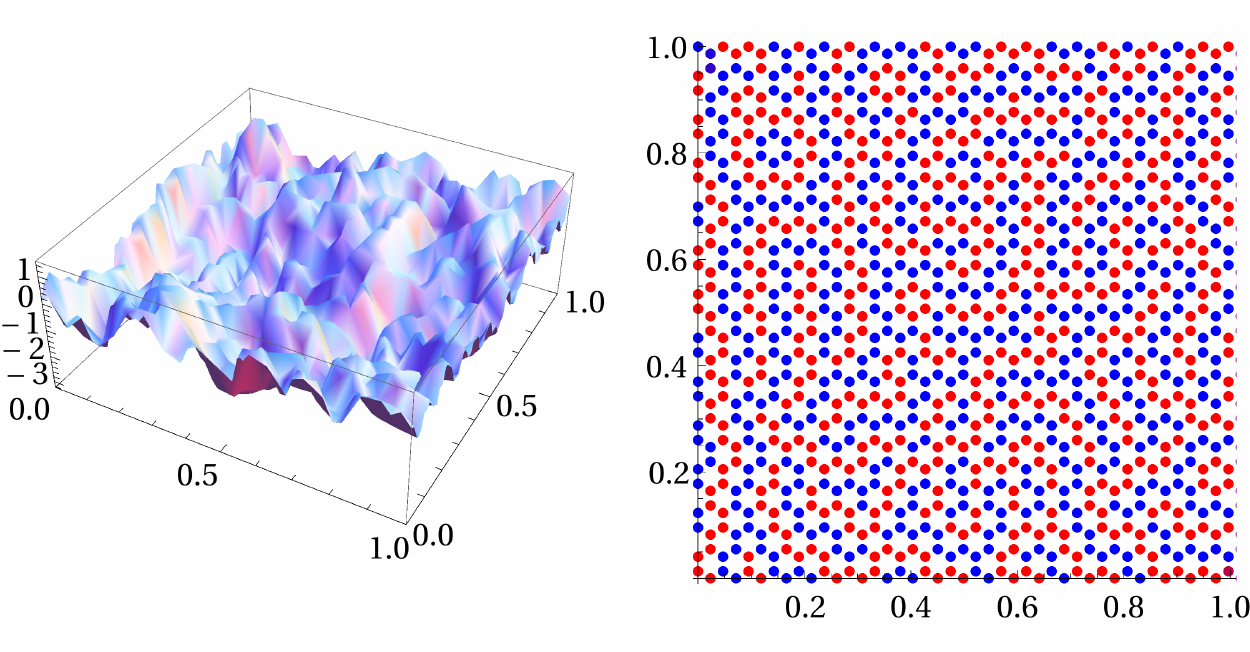}\\
H
\includegraphics[width=2.75in]{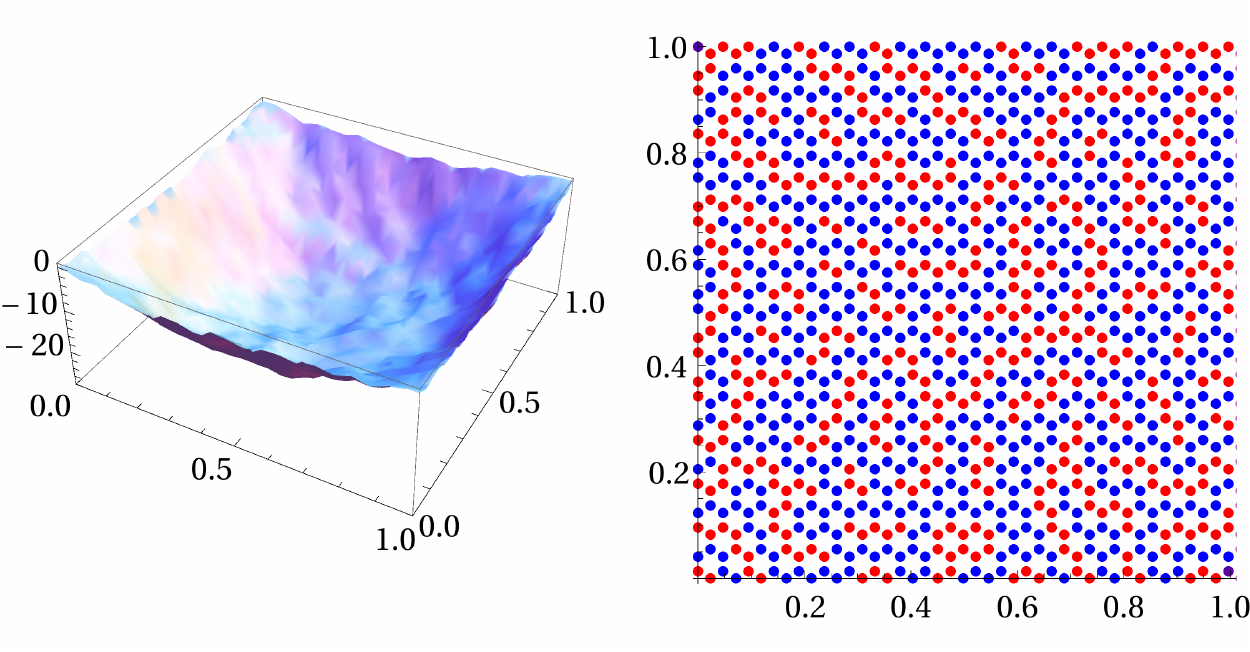}
\hfill
I
\includegraphics[width=2.75in]{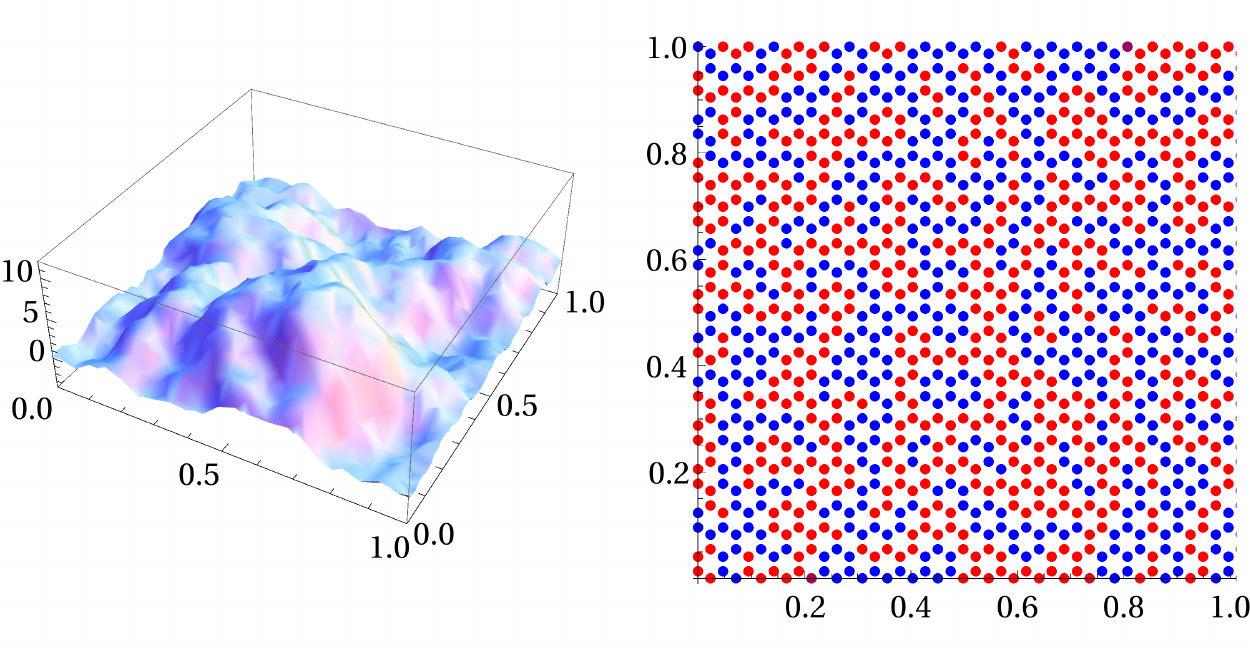}\\
J
\includegraphics[width=2.75in]{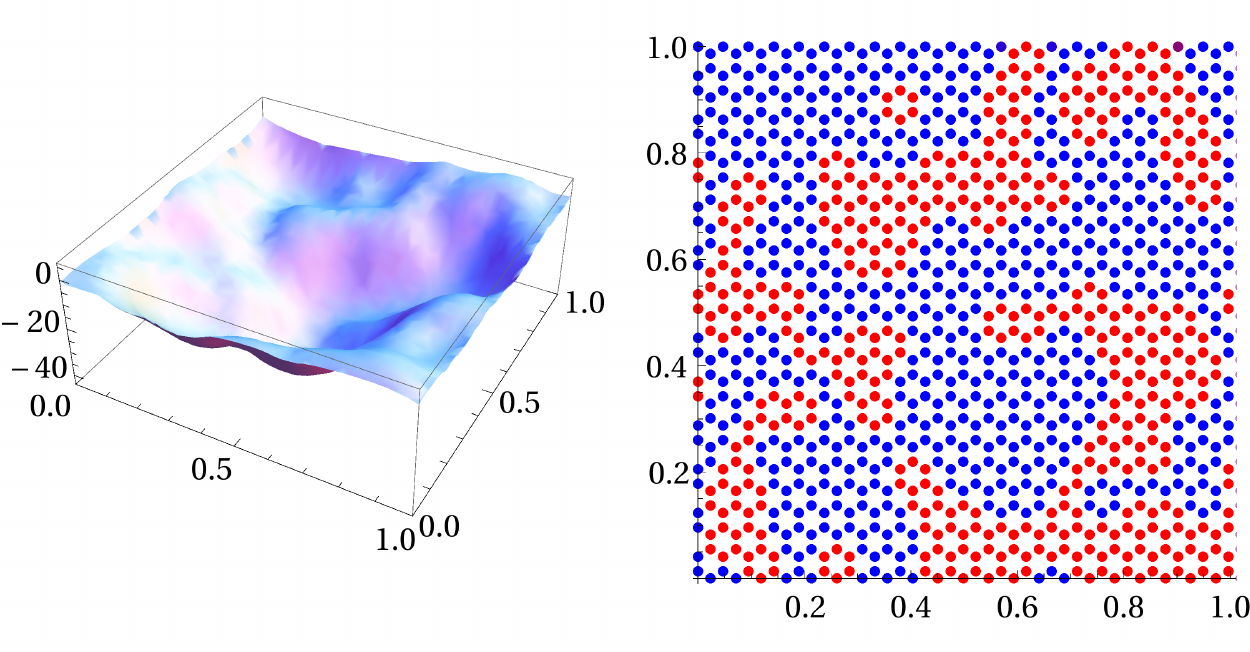}
\caption{\label{FG} Final configuration of the plane and the spins
  (red for spin up and blue for spin down) for different values of
  $\kappa$ and $\lambda$, corresponding to different regions of the
  phase diagram in Fig.~\ref{diag}. From top to bottom and left to
  right: Region $A$, $\kappa/\theta=-10$ and $\lambda/\theta=2$, Region $B$,
  $\kappa/\theta=0$ and $\lambda/\theta=10$, Region $C$, $\kappa/\theta=40$ and
  $\lambda/\theta=3$, Region $E$, $\kappa/\theta=40$ and $\lambda/\theta=27$, Region $F$,
  $\kappa/\theta=0$ and $\lambda/\theta=24$, Region $G$, $\kappa/\theta=-25$ and
  $\lambda/\theta=24$, Region $H$, $\kappa/\theta=-60$ and $\lambda/\theta=30$, Region $I$,
  $\kappa/\theta=-73$ and $\lambda/\theta=27$, Region $J$, $\kappa/\theta=-91$ and
  $\lambda/\theta=3$.  }
\end{figure}

\newpage

\begin{thebibliography}{99}
\bibitem{sci04nov}  Novoselov K S,  Geim A K,  Morozov S V,  Jiang D, Y. Zhang Y,  Dubonos S V,  Grigorieva I V and  Firsov A A, 2004  Science {\bf 306} 666 

\bibitem{PNA05nov}  Novoselov K S, Jiang D, Schedin F,  Booth T J,  Khotkevich V V,  Morozov S V and  Geim A K, 2005 Proc. Natl. Acad. Sci. USA {\bf 102} 10451

\bibitem{RMP09cas}  Castro Neto A H, Guinea  F,  Peres N M R,  Novoselov K S,
and  Geim A K, 2009 Rev. Mod. Phys. {\bf 81} 109

\bibitem{nature07mey} Meyer J C, Geim A K, Katsnelson M I, Novoselov K S, Booth T J and  Roth S, 2007 Nature {\bf 446} 60

\bibitem{pss09ban} Bangert U, Gass M H, Bleloch A L, Nair R R, and Geim A K, 2009 Physica status solidi (a) {\bf 206} 1117

\bibitem{PRB08gui} Guinea F, Horovitz B and Le Doussal P, 2008 Phys. Rev. B {\bf 77} 205421

\bibitem{PTRS08kat} Katsnelson M I and Geim A K, 2008 Phil. Trans. R. Soc. A {\bf 366} 195

\bibitem{ASC11mao} Mao Y, and Wang W L, Wei D, Kaxiras E, and Sodroski J G, 2011 ACS Nano. {\bf
    5} 1395

\bibitem{nat07fas} Fasolino A, Los J H and Katsnelson M I, 2007 Nature 
Materials {\bf 6} 858


\bibitem{PRB07abe} Abedpour N, Neek-Amal M, Asgari R, Shahbazi F, Nafari N and  Tabar M R, 2007 Phys. Rev. B {\bf 76} 195407 

\bibitem{EPL08Eun}  Kim E A and  Castro Neto A H, 2008 Europhys. Lett. {\bf 84} 57007

\bibitem{PRB09gaz}  Gazit D, 2009 Phys. Rev. B {\bf 80} 161406(R) 

\bibitem{PRL11san}  San-Jose P, Gonz\'alez J and Guinea F, 2011 Phys. Rev. Lett. {\bf 106} 045502 

\bibitem{PRB14gon}  Gonz\'alez J, 2014 Phys. Rev. B {\bf 90} 165402

\bibitem{PRB14gui}  Guinea F, Le Doussal P and Wiese K J,
 2014 Phys. Rev. B {\bf 89} 125428 
                                
\bibitem{PRB14amo}
Amorim B, Rold\'an R, Cappelluti E, Fasolino A, Guinea F and Katsnelson MI, 
2014 Phys. Rev. B {\bf 89} 224307 

\bibitem{zan12} 
R. Zan, C. Muryn, U. Bangert, P. Mattocks, P. Wincott, D. Vaughan, X. Li, L. Colombo, R.S. Ruoff, B. Hamilton and K.S. Novoselov, 
2012 Nanoscale {\bf 4}, 3065

\bibitem{sch15}
J.K. Schoelz, P. Xu, V. Meunier, P. Kumar, M. Neek-Amal, P. M. Thibado, and F. M. Peeters, 
2015 Phys. Rev. B {\bf 91}, 045413 

\bibitem{nl12oha} O'Hare A, Kursmartsev F V and Kugel K I, 2012 Nano Lett. {\bf
    12} 1045

\bibitem{pre12bon} Bonilla L L, Carpio A, Prados A and Rosales R R, 2012 Phys. Rev. E {\bf 85} 031125 

\bibitem{jsm12bon} Bonilla L L and Carpio A, 2012 J. Stat. Mech.:
  Theor. Exp. P09015  

\bibitem{jstat10} Prados A, Bonilla L L and Carpio A, 2010 J. Stat. Mech.:
  Theor. Exp. P06016;   Bonilla L L, Prados A and Carpio A, 2010 J. Stat. Mech.:
  Theor. Exp. P09019

\bibitem{bc} We take the spins at the boundaries formally
    equal to zero because it is the simplest choice. Alternatively,
    this boundary condition may be understood as if there were no spin
    associated to the clamped displacements $u_{0}=u_{N+1}=0$. From
    either point of view, the term corresponding to the
    spin-oscillator and the spin-spin
    interactions in eq.~\eqref{H1} become
    $-f\sum_{j=1}^{N}u_{j}\sigma_{j}+J\sum_{j=1}^{N-1}\sigma_{j+1}\sigma_{j}$,
    which only involve the ``bulk'' sites $j=1,\ldots,N$.


\bibitem{Gl63} Glauber R J, 1963 J. Math. 
Phys. {\bf 4} 294

\bibitem{unpub} Ruiz-Garc{\'\i}a M, Bonilla L L, and Prados A, to be published.

\bibitem{prb08car} Carpio A and Bonilla L L, 2008 Phys. Rev. B {\bf 78} 085406

\bibitem{prb12bon86} Bonilla L L and Carpio A, 2012 Phys. Rev. B {\bf 86}  195402  

\bibitem{note} For both ferromagnetic and antiferromagnetic
  ordering, the next-nearest-neighbors of a given spin are parallel to it.


\bibitem{mezard} Mezard M, Parisi G and Virasoro M A, 1987 \textit{Spin glass theory and beyond}, (Singapore: World Scientific) 
 
\bibitem{hemmen} Heidelberg Colloquium on Glassy Dynamics, (Van Hemmen J L and Morgenstern I, ed.), 1983 Lect. N. Phys. {\bf 192} 203-233 

\bibitem{edw} Edwards S F, and Anderson P W, 1975 Journal of Physics F: Metal Physics {\bf 5} 965

\bibitem{ruder} Ruderman R A and Kittel C, 1954 Physical Review {\bf 96} 99

\bibitem{binder} Binder K and Young A P, 1986 Reviews of Modern
  Physics {\bf 58} 4 








\end{thebibliography}

\end{document}